# Nanometric precision distance metrology via chip-scale soliton microcombs

Yoon-Soo Jang,[1,†,*] Hao Liu,[1] Jinghui Yang,[1] Mingbin Yu,[2] Dim-Lee Kwong,[2] and Chee Wei Wong,[1*]

[1] Fang Lu Mesoscopic Optics and Quantum Electronics Laboratory, University of California, Los Angeles, CA 90095, USA.

[2] Institute of Microelectronics, Singapore 117685, Singapore.

[†] Present address: Division of Physical metrology, Korea Research Institute of Standards and Science (KRISS), 267 Gajeong-ro, Yuseong-gu, Daejeon, 34113, Republic of Korea.

* ysjang@ucla.edu; cheewei.wong@ucla.edu

**Laser interferometry serves a fundamental role in science and technology, assisting precision metrology and dimensional length measurement. During the past decade, laser frequency combs – a coherent optical-microwave frequency ruler over a broad spectral range with traceability to time-frequency standards – have contributed pivotal roles in laser dimensional metrology with ever-growing demands in measurement precision. Here we report spectrally-resolved laser dimensional metrology via a soliton frequency microcomb, with nanometric-scale precision. Spectral interferometry provides information on the optical time-of-flight signature, and the large free-spectral range and high-coherence of the microcomb enables tooth-resolved and high-visibility interferograms that can be directly readout with optical spectrum instrumentation. We employ a hybrid timing signal from comb-line homodyne interferometry and microcomb spectrally-resolved interferometry – all from the same spectral interferogram. Our combined soliton and homodyne architecture demonstrates a 3-nm repeatability achieved via homodyne interferometry, and over 1,000-seconds stability in the long-term precision metrology at the white noise limits.**

## Introduction

With length as one of seven fundamental physical quantities, the ability to precisely determine distance to a target is especially important such as in observations of gravitational-waves [1] and futuristic space missions of multiple satellite flying-formations [2]. With the current international system of units (SI) meter definition based on light vacuum path traveled in a time of 1/299,792,458 second [3,4], laser-based distance measurement plays a pivotal role to advance length metrology with increasing precision. Most laser interferometers is based on the single-wavelength, with interferometric phase measurement to achieve sub-wavelength precision [5]. Inherently the single-wavelength laser interferometer measures distance by accumulating a displacement from initial to target position, with the non-ambiguity range bounded at half the selected electromagnetic wavelength. To overcome this limitation, absolute distance measurement – which determines distance by a single operation – has been advanced in various platforms [6-8].

The advent of the frequency comb, which enables the whole optical frequency span to have traceability to well-defined frequency standards in the microwave or optical domains [9-11], brought about a breakthrough in absolute distance measurements [12]. The broad spectrum and ultrashort pulse of the frequency comb enable advanced laser distance metrology including dual-comb interferometry [6,13], synthetic-wavelength interferometry [14-16], spectrally-resolved interferometry [17-20], multi-wavelength interferometry [21-23], and cross-correlated time-of-flight measurements [7,24]. Recently chip-scale microcavities have contributed to progress in laser frequency combs [25,26] including the generation of different frequency microcombs [27-30], temporal solitons [31,32], integrated low-power microcombs [33], and optical frequency synthesizers [34]. These microcombs enable applications such as low-noise microwave generation [35,36], optical communications [37], spectroscopy [38,39], and distance measurement at ≈100 nm precision [40,41].

Here we describe spectrally-resolved laser ranging via a soliton frequency microcomb, with precision length metrology at the few nanometers scale. A single microcomb is utilized, of which the spectrally-resolved interferometry (SRI) of the measurement and reference pulses engraves information on the optical time-of-flight. With the large (88.5-GHz) free-spectral range and high-coherence of our selected frequency microcomb, we directly readout the tooth-resolved and high-visibility interferogram via optical spectrum analyzers. We utilize a dual-pumping technique to stably generate the soliton mode-locking in a planar-waveguide $Si_3N_4$ microresonator. We describe the time-of-flight signal reconstruction via the integrated platform of comb-line homodyne interferometry, microcomb and background amplified spontaneous emission spectrally-resolved interferometry, from the same spectral interferogram. The comb-line homodyne interferometry is unwrapped with the relatively-coarse microcomb, which is itself unwrapped with the low-coherence amplified spontaneous emission interferometry in the comb-background to achieve a 3-nm precision over a 23-mm non-ambiguity range. We sample the long-term distance metrology over 1,000-seconds stability and an Allan deviation up to 100-seconds, with a 3-nm measurement repeatability achieved from homodyne interferometry. Our soliton and homodyne single-microcomb architecture is at the long-term white noise bounds and, for this focus on short-distance high-precision, even comparable to a few-Hz stabilized fiber frequency comb for reference. We further demonstrate measurement linearity in example positional calibration and referenced against a three-dimensional (3D) precision gauge block for principle demonstration.

**Measurement concept of soliton microcomb ranging by spectrally-resolved interferometry**

Figure 1 shows our laser dimensional measurement concept with the soliton frequency microcomb. The time-delayed measurement pulse is described by relative phase delay $\phi(v)$ (= $2\pi v \times \tau_{TOF}$) to the reference pulse, with $v$ the optical carrier and $\tau_{TOF}$ the measurement-reference time-of-flight delay. With $s(v)$ the pulse spectrum, two separated pulses generate a frequency interference pattern as $i(v) = s(v) [1 + \cos \phi(v)]$, engraving the time delay with $1/\tau_{TOF}$ period (Fig. 1 right panel) with target distance $L$ determination from $2n_{air}L = (c_o\tau_{TOF})$ where $c_o$ is the vacuum speed of light and $n_{air}$ the medium refractive index. The time delay $\tau_{TOF}$ is directly determined by the peak position of the $i(v)$ Fourier-transform, expressed as $I(\tau) = FT\{i(v)\} = S(\tau) \otimes [\delta(\tau+\tau_{TOF})/2 + \delta(\tau) + \delta(\tau-\tau_{TOF})/2]$, where $\delta(\tau)$ is the Dirac delta function, $S(\tau)$ the $s(v)$ Fourier transform, and $\tau$ the time delay. Since $s(v)$ is a real function, its Fourier transform $S(\tau)$ is symmetric about $\tau = 0$ and repeated every $\tau = \tau_{pp}$ (=$1/f_r$), where $f_r$ is the repetition-rate and $\tau_{pp}$ the pulse-to-pulse temporal separation.

Since the $\tau_{TOF}$ peak is symmetrical to $\tau_{pp}/2$, the measured $\tau_{TOF}$ folds at $\tau_{pp}/2$ and the measured distance has a triangle-shaped profile with increasing target distance [17]. Thus, the target distance is expressed as $2n_{air}L = c_o(m \times \tau_{pp}/2 + \tau_{TOF})$ for even $m$, or $2n_{air}L = c_o\{(m+1) \times \tau_{pp}/2 - \tau_{TOF}\}$ for odd $m$, where $m$ is an integer [18]. In general, the calculated $\tau_{TOF}$ from $S(\tau)$ peak detection is not precisely determined and is limited by the $s(v)$ bandwidth. We use curve-fitting algorithm for precision peak detection and homodyne interferometry toward nanometer-level precision, with further details in Supplementary Materials S2. Note that the non-ambiguity range ($L_{NAR}$) is determined from $f_r$ [$L_{NAR} = c/(4f_r)$] which corresponds to 850-μm ($f_r$=88.5 GHz). For further precise measurement, we use the optical carrier phase from the inverse Fourier transformation of $S(\tau)$. Then the target distance can be defined as $2n_{air}L = c_0/v(M_{homodyne} + \phi(v))$, where $M_{homodyne}$ is the integer of the homodyne interferometry.

Figure 2 depicts the setup for the microcomb-based dimensional metrology. A dissipative single-soliton is generated in a planar $Si_3N_4$ microresonator, with loaded quality-factor $Q$ of $1.77 \times 10^6$, free spectral range (FSR) of 88.5 GHz, and anomalous group velocity dispersion $\beta_2$ of -$3 \pm 1.1$ fs$^2$/mm at 1595-nm. The stable single-soliton microcomb mode-locking is achieved with a counter-propagating dual-driven technique (see Supplementary Materials S1). Fig. 2(b) shows the generated single-soliton microcomb, formed in the microresonator with Kerr nonlinearity. The soliton microcomb has a hyperbolic secant-square shape with a 1595-nm center wavelength and a 40-nm bandwidth.

Figure 2(a) shows the experimental setup for absolute distance measurement. A C-band section of the soliton comb is first amplified with an erbium-doped fiber amplifier up to 10 mW, centered at 192-THz with 2-THz bandwidth. A 50:50 beam splitter divides the soliton microcomb pulses into the reference and measurement arms for the interferometry and recombines upon the pulses return. The measurement mirror ($M_{Mea}$) is mounted on a motorized stage for translational motion. The recombined beam is collimated into a single-mode fiber, and sent into an optical spectrum analyzer with 50-pm resolution and 10-pm accuracy, a fractional value of $6.3 \times 10^{-6}$. An example resulting spectral interference pattern is shown in the blue plot of Fig. 2(c). Since the microcomb has a large 88.5-GHz repetition-rate, the comb tooth-resolved interferogram can be directly readout with optical spectrum analyzer. In contrast, conventional fiber frequency combs rely on Fabry-Perot etalon-based mode filtering or virtually-imaged phase array spectrometers for comb tooth-resolved spectrograms. In Fig. 2(c), the background gray spectrum is the amplified C-band section of the original soliton microcomb for reference.

**Absolute distance measurement by soliton microcomb-based spectrally-resolved interferometry**

To evaluate the measurement reliability, we measured a fixed distance over 1,000-seconds with an 1-second update rate. The air refractive index during measurement is assumed at 1.000247 using the empirical equation under standard air [42]. Since non-ambiguity range of microcomb-based SRI is limited by hundreds of micrometer, we extend the non-ambiguity range by introducing coarse measurement from ASE spectrum-based SRI [43,44]. Since the spectrometer resolution is 50-pm ($\delta v_{spectrometer}$ = 6.14-GHz at 1560-nm), the maximum measurable range of ASE spectrum-based SRI ($L_{MAX\_ASE}$) is 23.4-mm by relation of $L_{MAX\_ASE} = c_o/2n\delta v_{spectrometer}$. Further measurement range extension can be realized by introducing other coarse distance metrology [18,23]. As shown in Fig. 3(b), the target distance ($L_{Mea} = c_o\tau_{TOF}/2$) and non-ambiguity range ($L_{NAR} = c_o\tau_{pp}/4$) is determined from the reconstructed time-domain signal, based on Fourier transform of the interference pattern in the frequency domain. Fig. 3(b) shows the non-ambiguity range at 0.847424-mm. We see a large peak enhanced by summation of the ASE and microcomb spectrum, extending the measurement range.

We evaluate the measurement linearity by comparing with the encoder inside the motorized stage as shown in Fig. 3(c). The measurements revealed a peak-to-valley discrepancy of ± 2.56-μm. We note that the comparison is limited by the motorized stage due to its low accuracy (≈ ±5-μm) (see Supplementary Materials S6). For further comparison, we compare the measurements between microcomb SRI and homodyne interferometry. The peak-to-valley discrepancy is ± 293-nm with standard deviation (1σ) of 185-nm. We also evaluated the translation motion exceeding the non-ambiguity range of 850-μm as shown in Fig. 3(c) inset. For comparison, the fiber comb-based SRI result is also plotted in light cyan. Beyond the non-ambiguity range, our measurement and the encoder matches well within the encoder accuracy. We also measured a standardized gauge block cross-section with 3-mm height to validate the microcomb SRI for potential 3D surface measurement. The measurement repeatability taken over 5 consecutive measurements is determined to be 327-nm and 11.4-nm from microcomb SRI and homodyne interferometry respectively as the 1σ standard deviation (detailed in Supplementary Materials Section S7). These results demonstrate that the microcomb SRI has a good potential for length and positioning calibration such as length standards and high-precision axial positioning.

As shown in Fig. 4(a) and 4(b) up to 1,000-seconds, the measured distance from microcomb SRI is nearly constant without notable long-term drifts and has a standard deviation (1σ) of 81.6-nm. In contrast, the ASE spectrum-based SRI shows large fluctuations in the distance measurement due to its incoherence, but aids to extend the measurement range via non-periodicity in the time-domain. The ASE spectrum-based SRI measurement range is instead usually limited by the spectrometer optical resolution. An average value of the measured distance from microcomb-based SRI is found to be 8.197951-mm, and its accuracy is estimated to be 52-nm bounded by the optical spectrum analyzer. This accuracy can be enhanced by precisely measuring the repetition-rate $f_r$, instead of reading out solely the spectrum analyzer values. Measurement repeatability (in terms of Allan deviation) is calculated via the long-term measurement as shown in Fig. 4(b). As noted in Fig. 4(b), the measurement repeatability of microcomb-based SRI at 1-second (without averaging) is found to be 80-nm. The measurement repeatability gradually improves to 11-nm, with a measurement fitted relation of 80-nm × $\tau_{avg}^{-0.5}$. For longer averaging time more than 10-seconds, the measurement repeatability remains below 20-nm.

Homodyne interferometry provides a complementary approach to further improve the distance metrology precision at the nanometric level, since it employs the optical carrier frequency, instead of the pulse train envelope in microcomb ranging. Using multiple comb lines, our comb-based homodyne interferometry counts the optical carrier phase and its measured distance has a standard deviation (1σ) of 10.4-nm during the 1,000-seconds integration. We observed slowly-varying fluctuations (random walk), as shown in Fig. 4(a). (We note that for specific ranges, such as 900 to 1,000-seconds in this case, the standard deviation improves to 3.6-nm.) An average value of the measured distance from homodyne interferometry is found to be 8.197915-mm. As shown in Fig. 4(b), the measurement repeatability of homodyne interferometry at 1-second is found to be 2.85-nm which deteriorated to 6.62-nm at 100-seconds. The measurement repeatability of microcomb-based SRI and homodyne interferometry are overlapping at more than 100-seconds of averaging because it is perhaps bounded by slowly-varying fluctuations on the optical path delay due to measurement path thermal expansion, air refractive index variations by slowly-varying environmental drift, or long-term fluctuations of the measured spectrum. If we remove long-term drift using high-pass filtering, the measurement stability can be enhanced to sub-nm at 100-seconds averaging as shown Supplementary Materials Fig. S5. We compare our soliton microcomb stability measurements with a fully-stabilized fiber frequency comb reference in the SRI (detailed in Supplementary Materials Section S4). The measurement repeatability is well-matched each other, verifying that our measurement stability is not limited by the soliton microcomb. In addition, details of the intensity fluctuations on the measurement stability is further described in Supplementary Materials Section S3.

We have examined the scaling of the microcomb SRI with estimates of the microcomb stability (detailed in Figure S8). For distances smaller than 1-m, our precision limit is bounded by the measurement repeatability. For distances more than 1-m, the measurement precision will be bounded by the frequency stability of our free-running frequency microcomb which has been reported at the $10^{-9}$ level [45]. The scaling is square-root proportional with distance since for the longer distances the precision limit is bounded by the microcomb frequency instability ($\Delta f/f \sim \Delta L/L$). When locking the free-running comb to a Rb atomic clock, the frequency stability can be brought down to $10^{-12}$ at 100-seconds integration, further improving the long-distance precision of the single-soliton microcomb spectrally-resolved interferometry.

**Acknowledgments**

The authors acknowledge discussions with Ki-Nam Joo, Abhinav Kumar Vinod, Wenting Wang, Jinkang Lim, Ken Chih-Kong Yang, and Li-Yang Chen. The authors acknowledge support from the Office of Naval Research (N00014-16-1-2094), the Lawrence Livermore National Laboratory (contract B622827), and the National Science Foundation. Y.-S. J. designed and led the work. H. L. led the soliton generation. Y.-S. J. performed distance measurements. J. Y., M. Y. and D.-L. K. nanofabricated the microresonator. Y.-S. J. and C. W. W. performed the measured data analysis. All authors discussed the results. Y.-S. J. and C.W.W. prepared the manuscript.

**Figures and Tables**

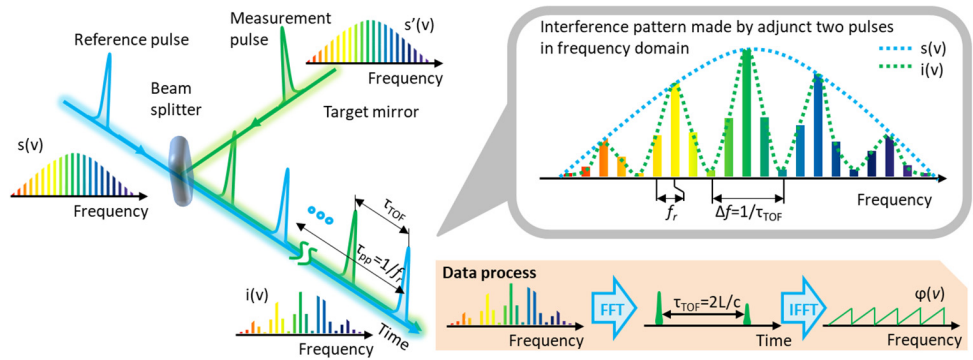

FIG. 1. Architectural approach of the spectrally-resolved ranging via soliton microcomb. Reference and measurement pulses of the soliton frequency comb, separated by $\tau_{TOF}$. The measurement pulse has a relative phase shift $\phi(v)$ [$= 2\pi v \times \tau_{TOF}$] to the reference pulse, and it makes an interference in every frequency mode of the soliton frequency comb. The information of $\tau_{TOF}$ is thus engraved on the interference pattern in the frequency domain, monitored via the spectrometer. The wide free-spectral range of frequency microcombs enables its comb-tooth resolved spectral interferogram to be directly readout by readily-available optical spectrum analyzers.

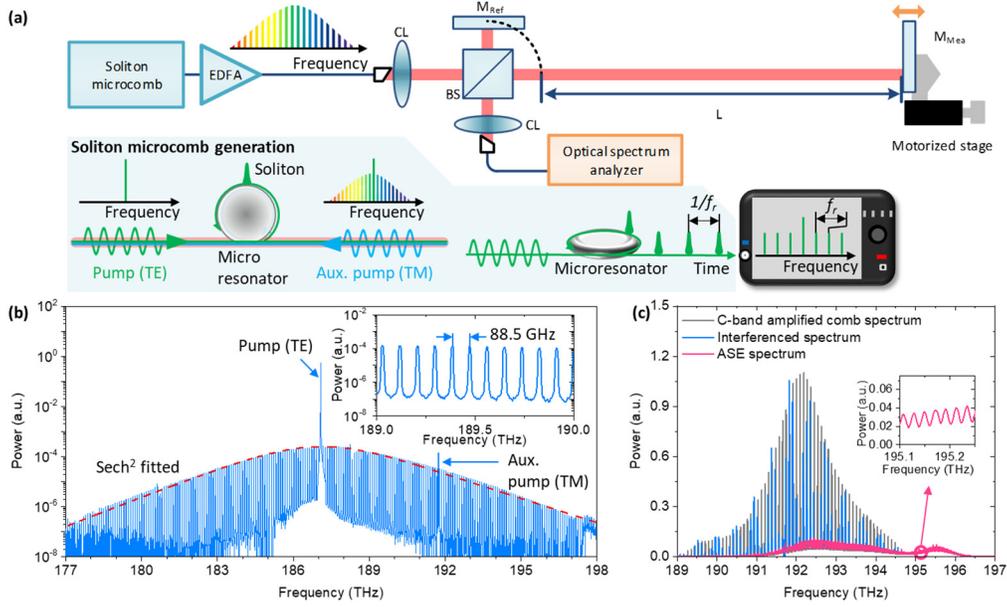

FIG. 2. Soliton microcomb-based precision dimensional metrology. (a) Soliton microcomb-based precision dimensional metrology via spectrally-resolved interferometry. BS: non-polarizing beam splitter, $M_{REF}$: reference mirror, $M_{MEA}$: measurement mirror, EDFA: erbium-doped fiber amplifier, CL: free-space collimator lens. Left inset describes schematic for the TE-TM dual-pumped soliton microcomb generation. (b) Example optical spectrum of the soliton microcomb from the high-$Q$ $Si_3N_4$ microcavity, with the hyperbolic secant-square spectrum. Inset: zoom-in illustration of the comb-tooth resolved spectrum. (c) Example measured high-coherence spectral interferogram (blue) from the reference and measurement pulses, along with the superimposed spectra of the C-band amplified soliton microcomb (gray). Pink line shows the amplified spontaneous emission (ASE) noise induced by the EDFA from the same spectral interferogram of the blue line. Inset: zoom-in illustration of the low-coherence ASE spectral interferogram with the low-visibility interference.

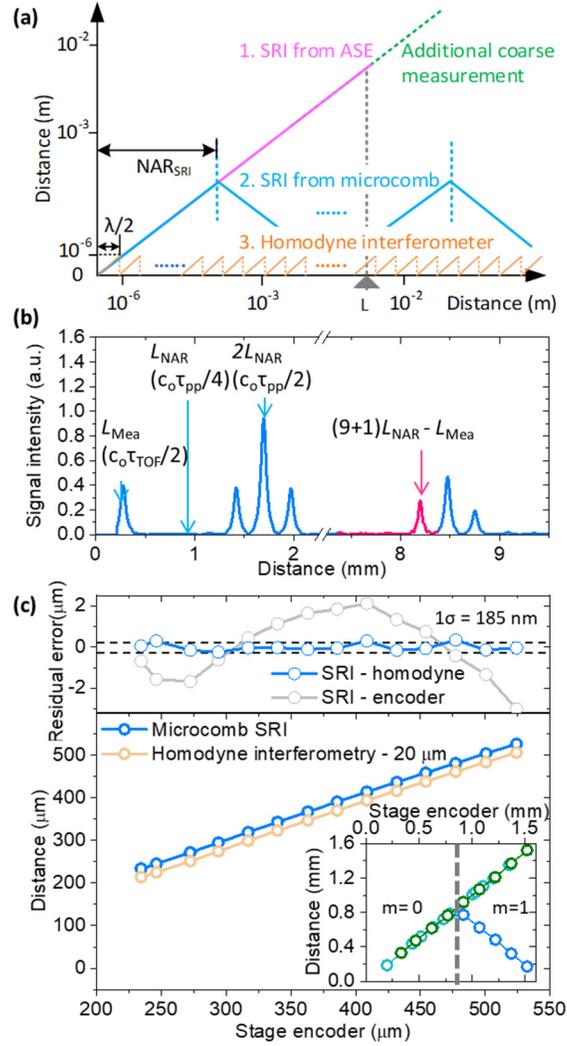

FIG. 3. Measurement linearity and distance measurement beyond the non-ambiguity range. (a) Non-ambiguity range extension by the combined platform of ASE-noise-based spectral interferometry, soliton microcomb spectral interferometry and homodyne interferometry. To determine integer $M$ of the homodyne interferometry, the coarse distance measurement from the microcomb (with $\lambda/2$, where $\lambda = c_o/\nu$.) is used. At the same time, to determine integer $m$ of the microcomb, the ASE SRI is used. (b) Time-domain signal reconstruction from the frequency domain interference. A typical signal-to-noise ratio of the time-domain signal is larger than 100. (c) Measurement linearity is evaluated by measuring the target distance with 25-μm incremental translation of a motorized stage. Left inset is the residual error comparing SRI, homodyne interferometry and the encoder. Right inset shows measured distance beyond the non-ambiguity range. The wrapped distance is unwrapped with calculated non-ambiguity in green color. For comparison, a distance measurement from fiber-comb based spectrally-resolved interferometry is also plotted in light green color.

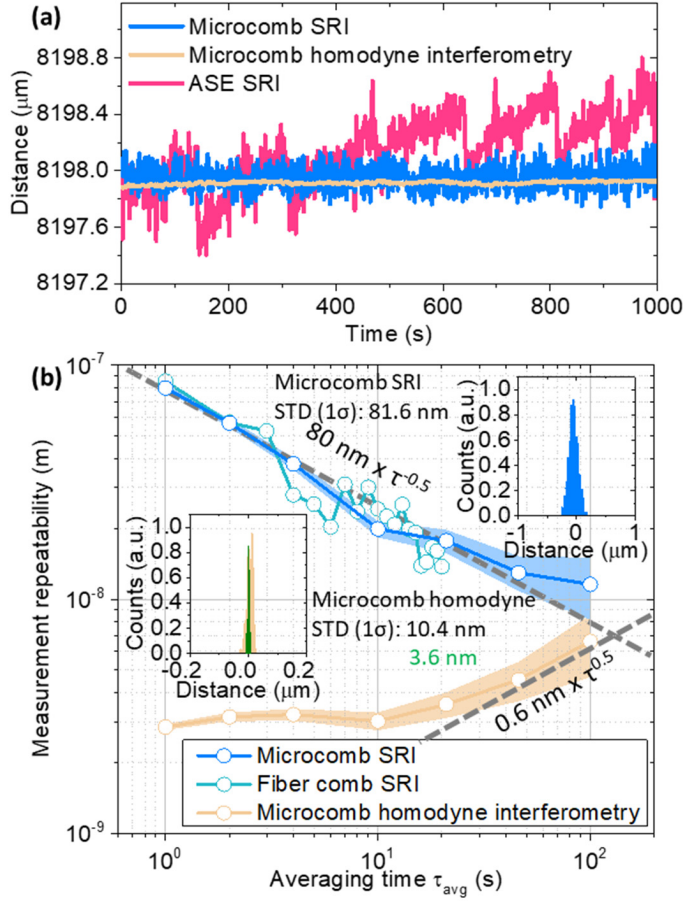

FIG. 4. Nanometer-scale precision distance measurement: reliability and repeatability evaluation. (a) Long-term distance metrology sampled over 1,000-seconds. (b) Measurement repeatability verification through Allan deviation of the long-term ranging data. 3-nm measurement repeatability is achieved from drift-compensated homodyne interferometry. The white noise limit is denoted by the dashed black line while flicker noise is not observed within our 100-seconds averaging time. Measurements from a few-Hz-stabilized fiber mode-locked laser frequency comb laser metrology are illustrated for comparison. Left inset: histogram distribution of microcomb spectral interferometry, with 1σ standard deviation of 81.6-nm at 1,000-seconds measurement. Right inset: histogram distribution of homodyne interferometry, with 1σ standard deviation of 10.4-nm at 1,000-seconds measurement. The 3.6-nm standard deviation is an example obtained from 900 to 1,000-seconds.

# Supplementary Materials for

## Nanometric precision distance metrology via chip-scale soliton microcombs

Yoon-Soo Jang,[1,†,*], Hao Liu,[1] Jinghui Yang,[1] Mingbin Yu,[2] Dim-Lee Kwong,[2] and Chee Wei Wong,[1*]

[1] Fang Lu Mesoscopic Optics and Quantum Electronics Laboratory, University of California, Los Angeles, CA 90095, USA.
[2] Institute of Microelectronics, Singapore 117685, Singapore.
[†] Present address: Division of Physical metrology, Korea Research Institute of Standards and Science (KRISS), 267 Gajeong-ro, Yuseong-gu, Daejeon, 34113, Republic of Korea.

* ysjang@ucla.edu; cheewei.wong@ucla.edu

**This Supplementary Materials consists of the following sections:**

**Section S1. Stable dual-pump generation of the single-soliton frequency microcomb**
**Section S2. Data processing for the distance metrology**
**Section S3. Characterization of intensity fluctuations on the distance metrology**
**Section S4. Reference against a stabilized mode-locked fiber laser frequency comb**
**Section S5. Bounds on the ultimate measurement precision of homodyne interferometry**
**Section S6. Position calibration of motorized stage by soliton microcomb-based spectrally-resolved interferometry**
**Section S7. Gauge block measurement for 3D surface measurement**

### Section S1. Stable dual-pump generation of the single-soliton frequency microcomb
#### S1.A. Planar-waveguide $Si_3N_4$ microresonator frequency comb

The microresonator used for the single-soliton frequency comb generation is based on stoichiometric silicon nitride with 261 μm outer radius and 800 nm thickness. The loaded and intrinsic quality factors $Q$ are $1.77 \times 10^6$ and $3.4 \times 10^6$ respectively. The microresonator width is adiabatically varied from 1 to 4 μm to tune the dispersion and improve the single-mode mode-locking. Using swept-wavelength interferometry, the free spectral range (FSR) is found to be 88 GHz with an anomalous group velocity dispersion $β_2$ of -3 ± 1.1 $fs^2$/mm.

#### S1.B. Counter-propagating dual-pump technique

We set the pump laser (New Focus TLB-6700) at 1595 nm with 23 dBm power and TE polarization. The auxiliary laser (Santec TSL-510) is at 1565 nm with 33 dBm power and TM polarization. The pump laser is set to generate the single-soliton state with counter-clockwise propagation in microresonator, The auxiliary laser wavelength is set for effectively blue-detuning to thermally stabilize the planar waveguide Si3N4 microresonator with clockwise propagation, while the pump laser wavelength is set to generate the single-soliton state with counter-clockwise propagation in microresonator. The dual-driven counter-propagating technique separates the thermal hysteresis from the Kerr soliton dynamics [S1].

### S1.C. Single-soliton generation in microresonator

A single soliton is deterministically generated by cascaded four-wave mixing in the planar waveguide $Si_3N_4$ microresonator via cross-polarized dual-driven approach. A 33-dBm TM auxiliary laser centered at 1560 nm is sent into the $Si_3N_4$ microresonator, and slowly detuned into resonance. Then a 24-dBm TE pump is sent into the microresonator in the counter propagation direction [S1]. With the thermal hysteresis compensation via the TM auxiliary laser, a single soliton state is deterministically generated by tuning the TE pump wavelength to the effective red-detuning side of the pump cavity resonance.

## Section S2. Data processing for the distance metrology
### S2.A. Fundamental minimum and maximum measurement range

To determine the distance, the reference and measurement pulse should be separated in the time domain. The minimum measurable distance ($L_{min}$) is determined by pulse duration used in the distance measurement. $L_{min}$ can be expressed as $L_{min} = c_o/(2\Delta v)$, where $\Delta v$ is a spectrum bandwidth. In our case, $L_{min}$ is estimated to be 30 μm considering 5 THz spectrum bandwidth of soliton microcomb. The fundamental maximum measurable distance ($L_{max}$) is upper-bounded by the coherence length of the light source and can be expressed as $L_{max} = c_o/(2\delta v)$, where $\delta v$ is the linewidth of the light source. In our case, the $L_{max}$ limit is estimated to be 1 km considering the 150 kHz linewidth of soliton microcomb.

### S2.B. Nonlinear curve fitting for precise peak detection

To precisely determine the peak position $\tau_{TOF}$ in time domain, we implement polynomial curve fitting near peak position as $I(\tau) = A\tau^2 + B\tau + C$. Data points for curve fitting are symmetrically chosen with 3 or 5 points around the peak position. The peak position is determined when its first derivative is equal to zero as $dI(\tau)/d\tau = 2A\tau + B = 0$. Thus the peak position is simply determined from $\tau = -B/2A$ (detailed in below section).

### S2.C. High precision distance measurement by homodyne detection from microcomb spectral interferometry

Spectrally-resolved interferometry [S2] has been examined to understand the frequency microcomb coherence [S3-S6]. Multi-wavelength interference has also been examined for absolute distance metrology [S7, S8]. For our distance metrology based on the microcomb-enabled spectral resolved interferometry, Figure S1 shows further details on our data processing. Firstly, the interference pattern in frequency domain was recorded by optical spectrum analyzer (Yokogawa, AQ6370) with 8.6 THz bandwidth. The measured interference pattern ($i(v) = s(v) [1 + \cos \phi(v)]$) shows sinusoidal modulated shape with period of $1/\tau_{TOF}$ in frequency domain due to optical carrier frequency depended relative phase delay ($\phi(v) = 2\pi v \tau_{TOF}$). The frequency domain signal is converted into time domain ($I(\tau) = FT\{i(v)\} = S(\tau) \otimes [\delta(\tau+\tau_{TOF})/2 + \delta(\tau) + \delta(\tau-\tau_{TOF})/2]$) by Fourier transformation. To simply determine $\tau_{TOF}$, a position of maximum intensity can be chosen, however, its resolution is restricted by temporal resolution of Fourier transformation. It can be enhanced by zero-padding technique, however, it requires much computational time with increasing number of zero-padding points [S9, S10], and its effect is described in next Section. Alternatively, we have nonlinear curve fitting to finely detect peak position of $\tau_{TOF}$ as described in the Methods section of main text. However, such envelope peak detection-based distance metrology cannot support nanometric precision distance measurement. Consequently, to improve measurement precision, we use homodyne detection from microcomb spectral interferometry. A filtered time domain signal near $\tau_{TOF}$ is subsequently inverse-Fourier transformed back to the frequency domain as:

$$i'(v) = FT^{-1}\{ S(\tau) \otimes \delta(\tau-\tau_{TOF})/2\} = [s(v)\exp\{i(2\pi\tau_{TOF}v)\}]/2 = [s(v)\exp\{i\phi(v)\}]/2 \qquad (1)$$

where $i = (-1)^{1/2}$. This process allows the spectral phase $\phi(v)$ to be recovered. The spectral phase can be determined by the formula of $\phi(v) = \tan^{-1}[\text{Im}\{s'(v)\}/\text{Re}\{s'(v)\}]$. Then the target distance can be determined by $L = c/2v \times \{M_{Homodyne} + \phi(v)\}$, where $M_{Homodyne}$ is an integer value. Since peak detection-based distance measurement provides accurate distance to be enough to determine integer value $M_{Homodyne}$, we can use homodyne method with nanometric precision over long range [S11, S12].

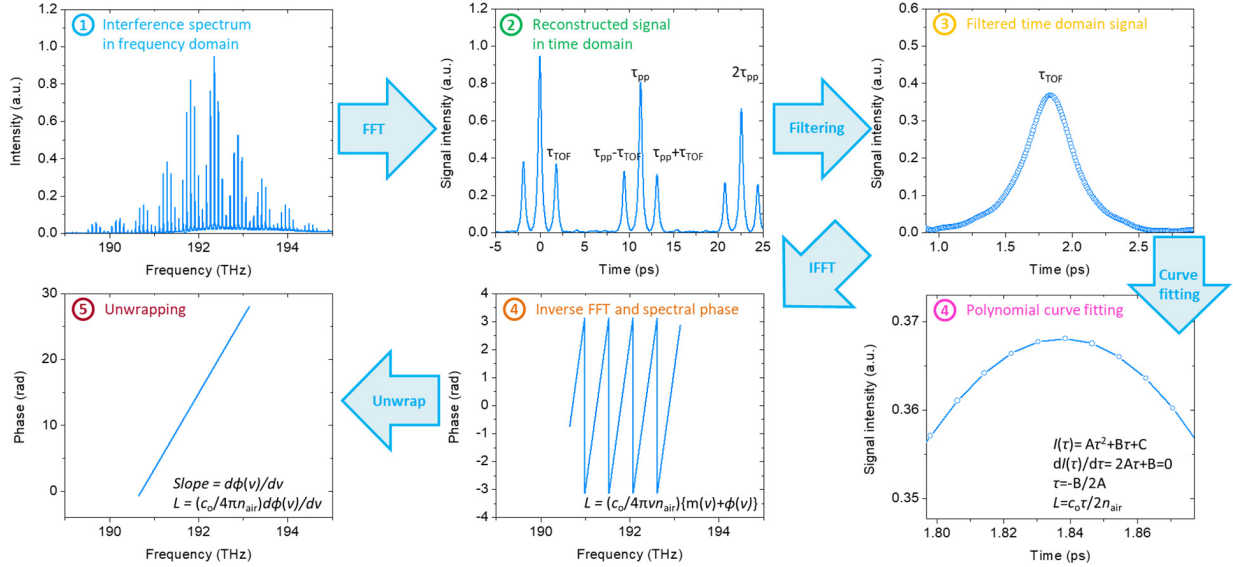

FIG. S1. Data process of spectrally-resolved interferometry. The target distance is determined by two steps. Firstly, time-domain peak detection is used to coarse measurement. Secondly, spectral phase extracted from inverse FFT spectrum from time-domain signal is used to homodyne interferometry for nanometric precision distance measurement. Alternatively, spectral phase slope can be also used to coarse measurement, since the first derivation of the phase delay for optical frequency $v$ ($d\phi(v)/dv = 2\pi\tau_{TOF}$) is proportional to $\tau_{TOF}$.

## S2.D. Comparison of peak detection method

Zero-padding technique makes Fourier-transformation data to be smoother [S13]. If we simply determine $\tau_{TOF}$ by reading out the position of maximum intensity, its resolution is limited by 115 fs temporal resolution considering 8.6 THz of spectral range of measured optical spectrum. It means that measured distance is digitized with 115 fs temporal interval as shown in Fig. S2 (a). In theory, the temporal resolution of Fourier transformation can be infinitely reduced, however, it comes with a large computational time to achieve high precision distance measurement. However, the measurement precision of nonlinear curve fitting method was found to be near 100 nm whether zero-padding is considered or not. Since the nonlinear curve fitting method do not need to zero-padding for improvement of measurement precision, we determined the distance using this approach. Alternatively, spectral phase slope [S14] or cross-correlation methods [S15] can be also considered for high-precision peak detection.

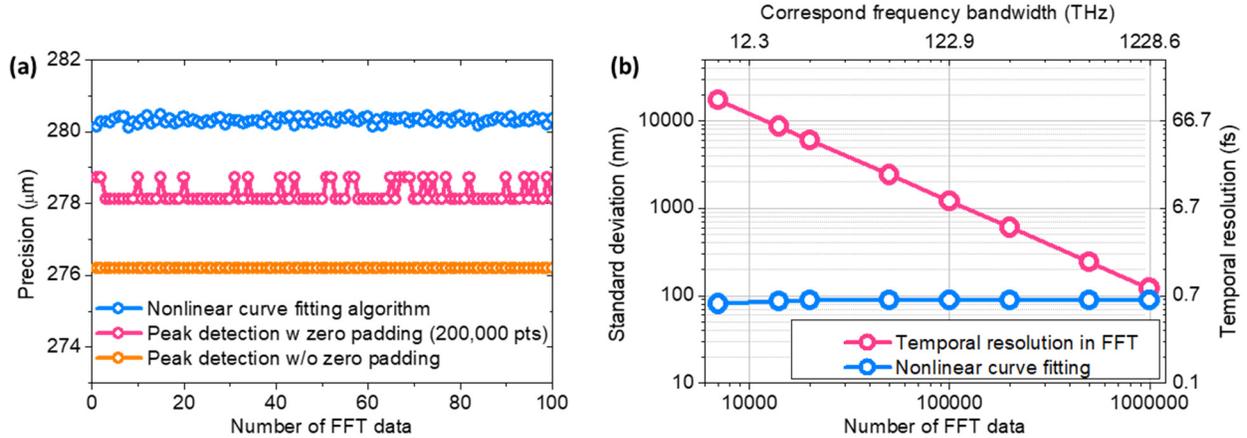

FIG. S2. Precision comparison between peak detection and nonlinear curve fitting method. (a) Time trace of peak detection and nonlinear curve fitting method. For peak detection method, two case (with zero padding and without zero padding) is plotted. For nonlinear curve fitting method, zero padding is not considered. Each result is shifted about 2 μm for comparison. (b) Measurement precision versus number of FFT data for zero padding.

**Section S3. Characterization of intensity fluctuations on the distance metrology**

Intensity fluctuations during the measurement deteriorates the interference pattern in frequency domain, which worsens the measurement precision. To verify this influence, we measure the distance with low and high intensity fluctuation state of soliton microcomb and have numerical simulation. From the interference pattern recorded by optical spectrum analyzer, one of comb line is used to monitor the intensity fluctuation. We investigate for two cases and the measurement results plotted with orange circle. For 2% of intensity fluctuation, the standard deviation value of measured distance is found to be 81.6 nm as shown in Figure S3. On the other hand, the standard deviation of measured distance is found to be 469 nm when intensity fluctuation is about 36%. To simulate this situation, one of interference pattern in frequency domain is used and its intensity is modulated by random fluctuation with range of 1% to 50%. Numerical simulation results is plotted in green color and it is quite well-matched with the experimental data. Note that the intensity fluctuations could be generated from the optical spectrum analyzer, the light source itself, polarization variation in the long fiber delay line, and also actual distance variations during measurement.

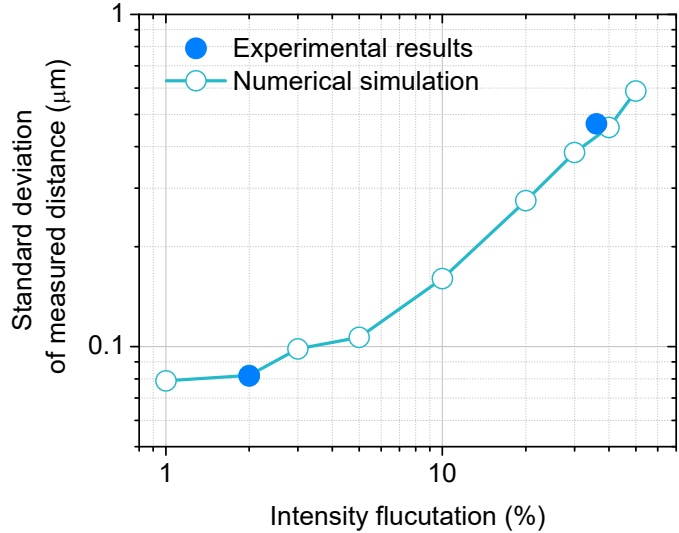

FIG. S3. Intensity fluctuations effects on the measurement precision. Green color denotes numerical simulation results of intensity fluctuation induced measurement precision variation. Blue dot denotes two examples of the experimental results.

**Section S4. Reference against a stabilized mode-locked fiber laser frequency comb**

A 250 MHz fiber comb (Menlo Systems) stabilized to 1 Hz laser with $10^{-15}$ fractional frequency stability (Stable Laser Systems) is used to verify our spectrally-resolved interferometry for laser ranging metrology [S16, S17]. The fiber comb has optical power of 10 mW and 1560 nm central wavelength. Since the spectrometer cannot resolve the interference pattern when its period is smaller than the resolution of the spectrometer, the measurement range of the fiber comb-based SRI is limited by the resolution of the spectrometer [S18]. For this reason, the target distance is fixed near 6 mm. The same interferometer and data processing are used for fiber comb-based SRI. Fig. S4 shows the measurement results of fiber comb-based spectral resolved interferometer. During the measurement of 100 seconds, the measured distance was nearly constant without any notable drift. The measurement repeatability is found to be 85.5 nm (24.5 nm) at averaging time of 1 second (10 seconds). The measurement repeatability of fiber comb-based SRI and soliton microcomb based SRI is almost identical. It means that the noise of soliton microcomb in our approach does not significantly contribute to the measurement repeatability.

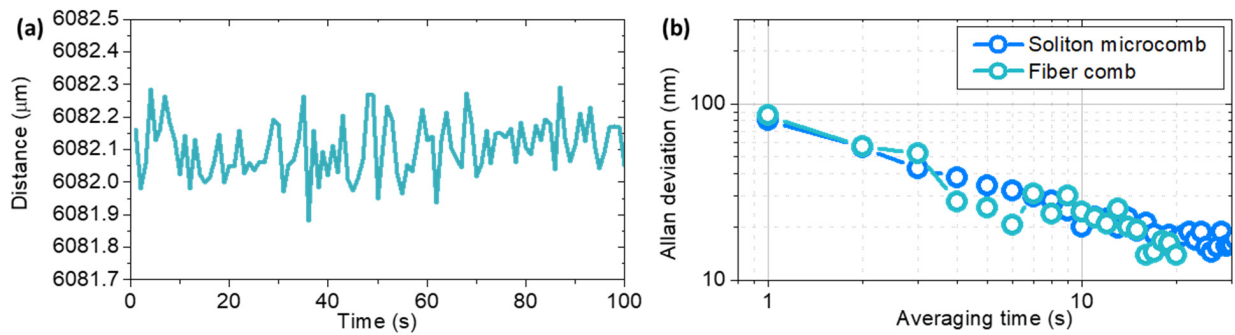

FIG. S4. Measurement result of fiber comb-based SRI. (a) Time trace of fiber comb-based SRI during 100 seconds. (b) Measurement precision in terms of Allan deviation.

**Section S5. Bounds on the measurement precision of homodyne interferometry.**

We found our measurement repeatability of the homodyne interferometry seems to be limited by environmental long-term drift including drift of refractive index of air and thermal expansion of the target distance. To evaluate ultimate measurement precision regardless of the long-term drift, we use 0.1 Hz high pass filter to minimize long-term drift effects on the precision. Fig. S5 shows comparison between raw data and high pass filtered data. For the high pass filtered case, a standard deviation (1σ) is improved to 3.2 nm and slowly-varying fluctuation disappears. If we assume that target is ideally fixed without long-term drift, measurement stability can be improved to be 0.1 nm at 100 seconds averaging time. Such measurement stability is close to commercial HeNe laser interferometry [S19].

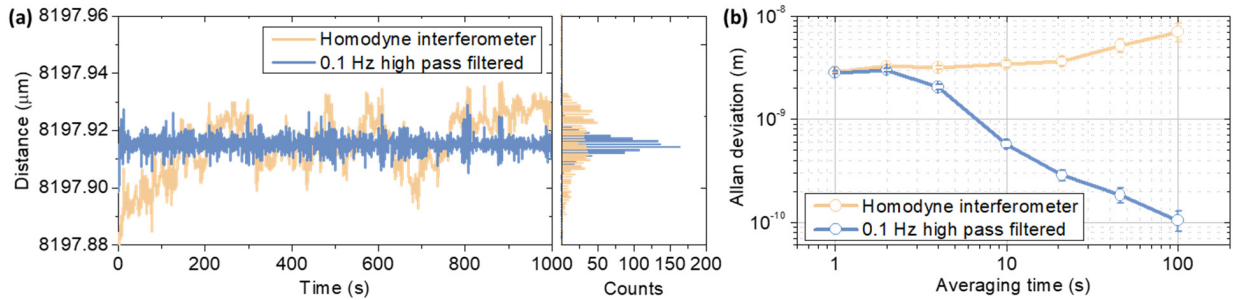

FIG. S5. Evaluation of measurement repeatability of homodyne interferometry. (a) Time trace of homodyne interferometry during 1,000 seconds with raw data marked in yellow color. Its 0.1 Hz high pass filtered data is also plotted with gray color. Right inset shows those histogram (b) Measurement precision in terms of Allan deviation, with 0.1 Hz high-pass filtering to remove the long-term drift. Sub-nm measurement stability at 100 seconds averaging can be observed.

**Section S6. Position calibration of motorized stage by soliton microcomb-based spectrally-resolved interferometry**

To verify the linearity of soliton microcomb based SRI, a motorized stage (New Focus MFN25) is used for comparison measurement. However, its low accuracy for long-stroke translation makes measurement range for the linearity test to be limited less than 150 μm. According to data sheet from manufacturer, on axis accuracy of the motorized stage is 10 μm. To calibrate position error of motorized stage, we compare the stage encoder value and measured distance by fiber comb-based SRI and soliton microcomb based SRI. Our measurement found that on-axis accuracy of the motorized stage is about ± 6 μm with cycle of 500 μm. This sinusoidal shaped cyclic error might be caused by mechanical structure of the motorized stage. We also found that linearity of the motorized stage is well maintained within 1 μm level at short range of 150 μm. Hence we choose this part for linearity evaluation of soliton microcomb based distance measurement.

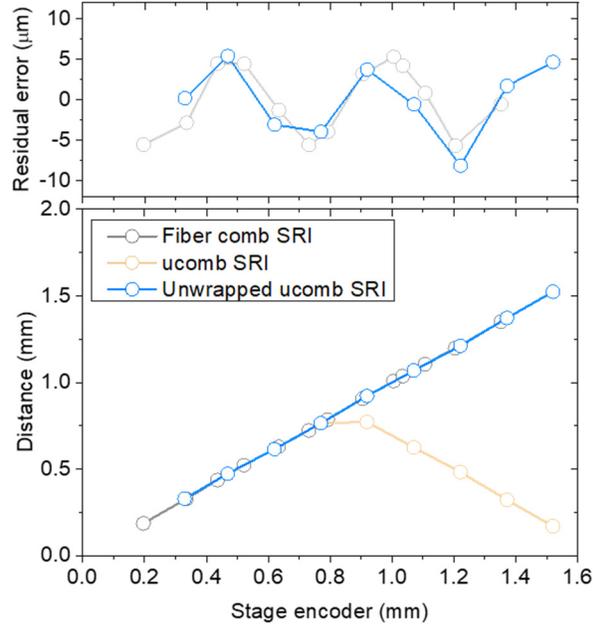

FIG. S6. Evaluating the accuracy of motorized stage by fiber comb and soliton microcomb based. Lower panel shows distance measurement results from fiber comb and microcomb based spectral resolved interferometry versus the motorized stage encoder. Both independent measurements of fiber comb and microcomb based spectral resolved interferometry show positioning error of the motorized stage encoder about ± 6 μm with cycle of 500 μm.

**Section S7. Position calibration of motorized stage and measurement range dependent imprecision by soliton microcomb-based spectrally-resolved interferometry**

To validate the microcomb SRI for potential 3D surface measurement, we measured a cross-section of a standardized gauge block, used for practical length metrology in 3D surface measurements and industry standards. We replaced the reference mirror in the interferometer part with the gauge block to measure the cross-section of the 3 mm height gauge block (Starrett RCM, 3.0 Al) that has a 300 nm uncertainty. The reference beam is made with a 4% Fresnel reflection from the end of FC/PC fiber ferrule. The transmitted beam is reflected from the target surface and sent to the optical spectrum analyzer along with the reference beam. The gauge block is mounted on a flat mirror and the stage made an on-axis translation with 1.27 mm (0.05 inch) steps as shown in Fig. S7 (a). The distance at each step is recorded with 5 data points. The gauge block height is determined by the difference of the absolute distances between mirror and gauge block surface, with the same empirical air refractive index of 1.000247 as noted above. The measured cross-section of the gauge block is shown in Fig. S7 (b) and (c). The height of gauge block was found to be 3.001237 mm and 3.001104 mm from the microcomb SRI and homodyne interferometry respectively. We also found a different slope height between the mirror (1.413μm/mm) and gauge block surface (-1.817μm/mm). A tilting (cosine) error from imperfect plane-to-plane alignment may introduce the measurement error of 1.237 μm. As shown in Fig. S7 (d), the measurement repeatability taken over 5 consecutive measurements is determined to be 327 nm and 11.4 nm from microcomb spectral resolved interferometer and homodyne interferometry, respectively as the 1σ standard deviation.

In the main text, we demonstrated the spectrally-resolved interferometry provides 3-nm precision with 23-mm non-ambiguity range in the free-running soliton frequency microcomb. With the proportional scaling to longer distances, the measurement precision will depend on the measurement range. In external field-operating scenarios, air refractive index has an ≈ $10^{-6}$ level fluctuation in uncontrolled environments and can be compensated to the $10^{-8}$ level with well-defined empirical estimates [S20] or two-color interferometry [S21]. If we assume the air refractive index and target vibrations are negligible, the measurement range dependent imprecision ($\Delta L$) can thus be estimated by $\Delta L = [(3 \text{ nm})^2 + \{(\Delta f/f) \times L\}^2]^{1/2}$. Our spectrally-resolved interferometry approach with both the soliton and comb-line homodyne interferometry can support distance measurements up to a kilometer or more, since the maximum measurable range ($L_{max}$) is bounded by the comb coherence length.

Fig. S8 summarizes our combined metrology specifications and measurement repeatability, scaling as a function of measurement range.

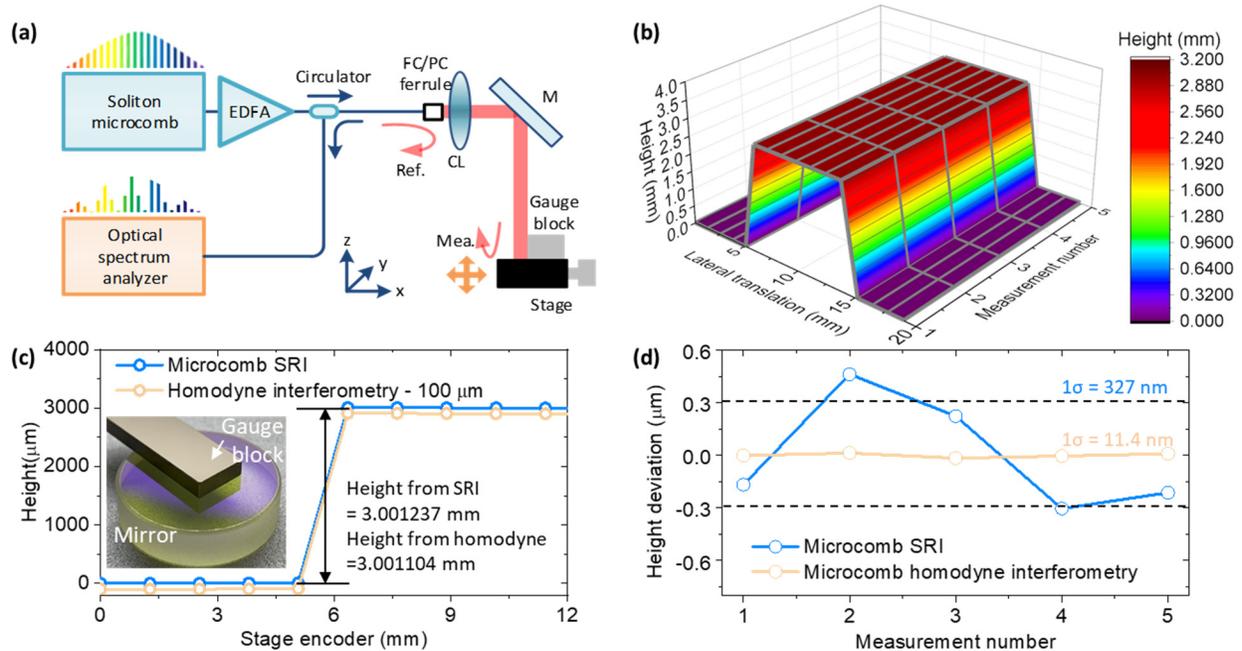

FIG. S7. Measurement of a reference gauge block cross-section via x-axis scanning. Metrology of a reference gauge block cross-section via x-axis scanning. (a) Measurement scheme for cross-section of gauge block with x-axis scanning stage. (b) Reconstructed cross-section of a gauge block. (c) The gauge block height is found to be 3.001237 mm and 3.001104 mm from soliton microcomb spectral interferometry and homodyne interferometry, matching well with reference specified height. (d) Measurement repeatability of the gauge block height.

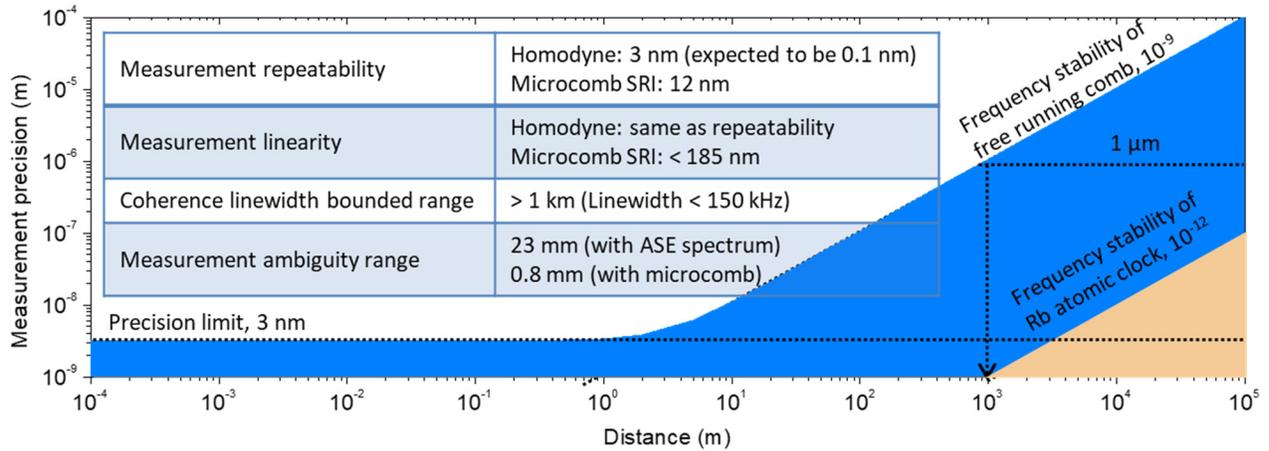

FIG. S8. Metrology specifications and measurement repeatability scaling as a function of measurement range.

**Supplementary References**